\documentclass[aps,pre,reprint,superscriptaddress]{revtex4-1}

\pdfoutput=1

\usepackage{graphicx}
\usepackage{amssymb}
\usepackage{amsmath}
\usepackage{xcolor}
\usepackage{pdfpages}
\usepackage{pgffor}
\usepackage{CJK}
\usepackage{multirow}
\usepackage{gensymb}

\makeatletter
\AtBeginDocument{\let\LS@rot\@undefined}
\makeatother

\begin{document}

\title{Inducing Stratification of Colloidal Mixtures with a Mixed Binary Solvent} 

\author{Binghan Liu}
\affiliation{Department of Physics, Center for Soft Matter and Biological Physics, and Macromolecules Innovation Institute, Virginia Tech, Blacksburg, Virginia 24061, USA}
\author{Gary S. Grest}
\affiliation{Sandia National Laboratories, Albuquerque, NM 87185, USA}
\author{Shengfeng Cheng}
\email{chengsf@vt.edu}
\affiliation{Department of Physics, Center for Soft Matter and Biological Physics, and Macromolecules Innovation Institute, Virginia Tech, Blacksburg, Virginia 24061, USA}
\affiliation{Department of Mechanical Engineering, Virginia Tech, Blacksburg, Virginia 24061, USA}

\begin{abstract}
Molecular dynamics simulations are used to demonstrate that a binary solvent can be used to stratify colloidal mixtures when the suspension is rapidly dried. The solvent consists of two components, one more volatile than the other. When evaporated at high rates, the more volatile component becomes depleted near the evaporation front and develops a negative concentration gradient from the bulk of the mixture to the liquid-vapor interface while the less volatile solvent is enriched in the same region and exhibit a positive concentration gradient. Such gradients can be used to drive a binary mixture of colloidal particles to stratify if one is preferentially attracted to the more volatile solvent and the other to the less volatile solvent. During solvent evaporation, the fraction of colloidal particles preferentially attracted to the less volatile solvent is enhanced at the evaporation front, whereas the colloidal particles having stronger attractions with the more volatile solvent are driven away from the interfacial region. As a result, the colloidal particles show a stratified distribution after drying, even if the two colloids have the same size.
\end{abstract}

\maketitle

\section{Introduction}\label{intro}

Evaporation is a ubiquitous nonequilibrium process in which molecules at the surface of a liquid escapes into the gas phase. \cite{Gerasimov2018Book} It plays many important roles. For example, evaporation is crucial for the ecosystem on the earth by being a critical step in the water cycle. When a material is dried, its liquid content is lost via evaporation and its wetness is reduced. Drying is therefore a useful technique for fabricating materials.\cite{Brinker2004, Zhang2010AdvMater,Zhou2017AdvMater,Zimmermann23DDT,Cai2021} Nanoparticles can be dispersed in a suitable solvent and assembled into various packed states (e.g., a colloidal crystal) after drying.\cite{Cai2021} Latex solutions are frequently applied to various surfaces and dried to create protective coating layers.\cite{Keddie1997,KeddieRouth2010} Evaporating solvent out of the mixtures of fillers and polymer solutions is often used for the fabrication of polymer-based composite materials.\cite{Jouault2014} Understanding evaporation and drying is thus of vital importance for a wide range of fields from environment to industry.\cite{Gerasimov2018Book,Routh2013,Thampi2023}

When a multi-component solution is dried, the flow induced by solvent evaporation competes with the diffusive motion of the solutes.\cite{Routh2013,Gelderblom2022,Thampi2023} A plethora of outcomes can be generated. One famous example is the coffee-ring effect,\cite{Deegan1997} where the capillary flow in an evaporating sessile droplet can transport solutes to the edge of the drop and deposit them at the contact line. Various instabilities can also occur in solution films or droplets undergoing solvent evaporation.\cite{Mitov1998,Bassou2009,Boulogne2015SM} In the case of latex films, self-stratification has been observed when the latex solution contains incompatible components as binders.\cite{Vink1996,Zahedi2018} These components can be dissolved in a common solvent but phase separate and stratify when the amount of solvent is reduced below a certain threshold during drying. The outcome of stratification is influenced by the surface tension of the binders, the kinetics of the drying process, and the presence of various additives. This type of stratification can be regarded primarily as an equilibrium effect.

Further studies show that self-stratification can occur in suspensions of compatible but polydisperse colloidal particles.\cite{Schulz2018} Such particles are not expected to segregate even in a dry state if the mixture is properly equilibrated. Therefore, such size-induced stratifying phenomena are nonequilibrium in nature and have recently attracted great interest.\cite{Zhou2017AdvMater,Schulz2018} By extending an earlier model developed by Routh and Zimmerman,\cite{Routh2004} Trueman et al.~\cite{Trueman2012JCIS} showed that stratified distribution of colloidal particles can be produced in a drying suspension of bidisperse colloidal mixtures, usually with the larger particles distributed closer to the evaporation front since they diffuse more slowly. They also presented experimental evidence with blended solutions of acrylic latices to support their numerical results.\cite{Trueman2012Langmuir} 
Fortini et al. later discovered the unexpected small-on-top stratification in the case of extremely fast drying rates.\cite{Fortini2016} This discovery has triggered a series of experimental,\cite{Martin-Fabiani2016,Makepeace2017,Cusola2018,LiuXiao2018,Carr2018,LiuWeiping2019,LiuWeiping2021,Schulz2021,Schulz2022,Palmer2023,Coureur2023,XiaoMing2019,LiuWendong2019,Hooiveld2023,Lee2020SM,Samanta2020} theoretical,\cite{Zhou2017,Sear2017,Sear2018,Rees-Zimmerman2021,HeBoshen2021,Tatsumi2020JCP} and computational \cite{Fortini2017,Howard2017,Howard2017b,Statt2017,Statt2018,Howard2020,Tang2018Langmuir,Tang2019Langmuir_control,Tang2019JCP_compare,Tang2022,Tatsumi2018,Park2022SM,Jeong2021} studies to reveal the underlying physical mechanism. It is now generally believed that the small-on-top stratification can be explained on the basis of diffusiophoresis, where colloidal particles are driven by the concentration gradient of other solutes (e.g., another type of particles) in the suspension.\cite{Brady2011,Sear2017,Sear2018,Rees-Zimmerman2021} As a polydisperse suspension undergoes rapid drying, colloidal particles are enriched at the evaporation front, creating concentration gradients. The diffusiophoretic drive on the larger particles by the concentration gradients of the smaller particles is stronger than that on the smaller particles by the concentration gradients of the larger ones. Such asymmetry, which was evident in the theoretical analysis of Zhou et al.\cite{Zhou2017}, pushes the larger particles out of the region near the evaporation front, thus creating a small-on-top stratified distribution of the particles.

In addition to serving as a platform for the exploration of rich nonequilibrium physics, suspensions that can self-stratify after drying can provide a foundation for the development of novel formulations to produce stratified coatings. Methods to control stratification outcomes are thus highly desirable. Martin-Fabiani et al.~\cite{Martin-Fabiani2016} designed colloidal coatings in which stratification can be switched on and off on demand by changing the pH value of the suspension. Via emulsion polymerization, they synthesized polymeric nanoparticles with a pH-responsive hairy layer, whose thickness is adjustable with pH. They blended such nanoparticles, which have smaller sizes, with larger particles whose sizes do not change significantly with pH to create a bidisperse suspension. As low pH values, the nanoparticle mixture has a large enough size ratio so that a small-on-top stratified coating is generated after rapid drying via the diffusiophoresis mechanism. However, as the pH value of the suspension is raised, the originally smaller nanoparticles swell substantially and the size contrast with the larger particles is reduced. As a result, stratification during evaporation is suppressed and a homogeneous coating layer is produced after drying. Tang et al.~\cite{Tang2019Langmuir_control} proposed that a temperature gradient can be used to tune stratification all the way from large-on-top to small-on-top, on the basis of the observation that thermophoresis of nanoparticles is also size-dependent.\cite{Tang2018Langmuir} Li et al. demonstrated that in a water-borne coating suspension containing silica nanoparticles, the distribution of silica nanoparticles can be controlled by fine-tuning their interaction with the latex particles in the coating.\cite{LiSiyu2023}

Despite the past efforts, more facile approaches to tune stratification that can be easily executed are still of great value, as all the methods proposed so far are difficult to implement on large scales. In this paper, we propose a new method based on a binary solvent mixture to induce and control stratification in a suspension containing a binary blend of nanoparticles. This method can be even used to stratify nanoparticles that have the same size and thus are not expected to stratify at all under most conditions.

Mixed solvents are frequently used for the synthesis and manipulation of colloidal and polymeric materials.\cite{Abdulla-Al-Mamun2009,Gordon2016,Guo2008PRL,Hoogenboom2008,Iyengar2016,Ye2012AM,Shi2019ACSAMI,OConnell2019} Solvent mixtures are already employed in latex formulations to create self-stratifying coatings.\cite{Abbasian2021} The evaporation behavior of liquid mixtures is also under active exploration because of its practical relevance to real systems and technologies where materials involved are typically multicomponent.\cite{Beverley2000,Jeong2021JMST,Aragon2023,Esposito2023,Tang2021PRL,Kim2018JFM} Interestingly, Song et al. showed that when the mixture of a volatile (methanol) and a much less volatile (1-butanol) component is evaporated, the two components develop opposite concentration gradients at the evaporation front, with 1-butanol enriched and methanol depleted near the interface.\cite{Song2016} It is interesting to explore if such inhomogeneity in solvent distribution developed during evaporation because of the contrasting volatilities of the mixed solvent components can be utilized to tune the distribution of colloidal particles in a drying film. In this paper, we present a proof-of-concept study based on molecular dynamics (MD) simulations to demonstrate that indeed a mixture of two liquid solvents with different volatilities can be used to even stratify nanoparticles that are identical except for their contrasting preference to the different solvent components.

This paper is organized as follows. In Sec.~\ref{sec:md}, the MD methodology of modeling liquid mixtures with Lennard-Jones monomers (the more volatile component) and dimers (the less volatile component) and suspensions of nanoparticles in such mixed solvents is described. In Sec.~\ref{sec:res}, the evaporation behavior of the liquid mixtures at various evaporation rates, bulk temperatures, and mass percent composition ratios is analyzed. Then the simulation results are presented on using such mixed solvents to induce nanoparticle stratification during drying. A brief summary is presented in Sec.~\ref{sec:conc}.

\section{Simulation Methodology}\label{sec:md}

A mixture of Lennard-Jones (LJ) monomer beads and dimers (i.e., two-bead molecules) is used as the model of a binary solvent. Non-bonded beads are point masses of mass $m$ interacting with each other through a standard LJ 12-6 potential,
\begin{equation}\label{eqn:lj_pot}
    U_\text{LJ}(r_{ij})=4\epsilon\left[\left(\frac{\sigma}{r_{ij}}\right)^{12} - \left(\frac{\sigma}{r_{ij}}\right)^{6}\right]~,
\end{equation}
where $\epsilon$ is the interaction strength, $\sigma$ is the unit of length, and $r_{ij}$ is the distance between two beads, $i$ and $j$. The LJ potential is truncated at $r_c = 3.0 \sigma$. All relevant units can then be expressed in terms of $m$, $\epsilon$, and $\sigma$. For example, the unit of time is $\tau = \sqrt{m\sigma^2 /\epsilon}$.

The two beads in a dimer are connected by the finitely extensible nonlinear elastic (FENE) potential,\cite{Kremer1990}
\begin{eqnarray}\label{eqn:FENE}
    U_\text{B}(r_{ij}) &=& -\frac{1}{2}KR_{o}^{2} \ln \left[1-\left( \frac{r_{ij}}{R_{o}} \right)^{2} \right] \nonumber \\
    & & + 4\epsilon \left[\left(\frac{\sigma}{r_{ij}}\right)^{12}-\left(\frac{\sigma}{r_{ij}}\right)^{6} \right] + \epsilon~,
\end{eqnarray}
where $K = 30 \epsilon/\sigma^2$, $R_o = 1.5 \sigma$, and $r_{ij}$ is again the inter-bead distance. The LJ term in $U_\text{B}(r_{ij})$ is truncated at $r_c=2^{1/6}\sigma$. The minimum of $U_\text{B}(r_{ij})$ is located at a separation $r_{ij} \simeq 0.96\sigma$.

\begin{figure}[htb]
    \centering
    \includegraphics[width=0.8\columnwidth]{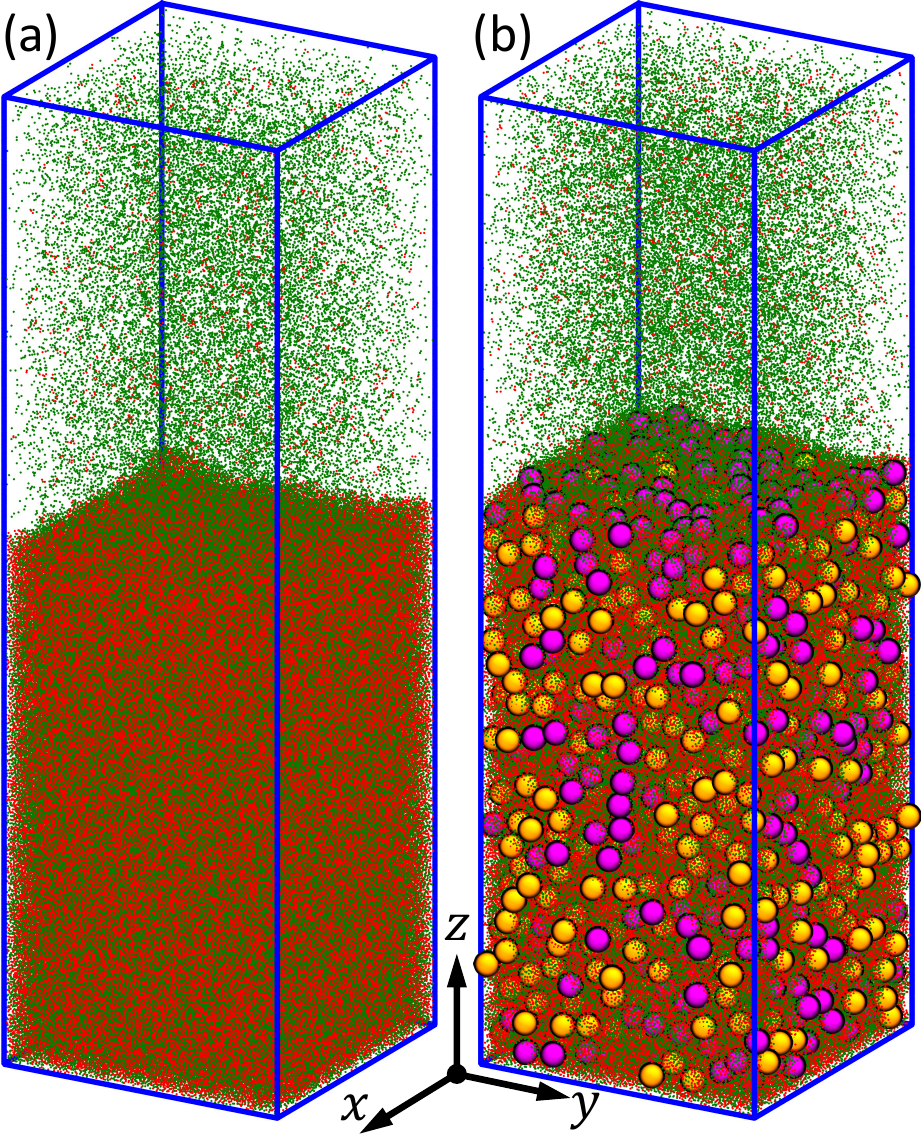}
    \caption{(a) Visualization of the CR$_1$ mixture of monomer (green) and dimer (red) beads at liquid-vapor equilibrium; (b) Visualization of an equilibrium state with a uniform dispersion of two types of particles (orange and purple) in the CR$_{1}$ mixture.}
    \label{fig:box}
\end{figure}

\begin{table}[htb]
    \centering
    \caption{Systems with different composition ratios (CRs)}
    \label{tab:systems}
    \begin{tabular}{| c | c | c | c | c |}
        \hline
        System & Total & Monomers & Dimers & CR \\ \hline
        CR$_{1}$ & 906512 & 452432 & 227040 & 1.0 \\ \hline
        CR$_{2}$ & 906498 & 604314 & 151092 & 2.0 \\ \hline
        CR$_{0.5}$ & 906498 & 302184 & 302157 & 0.5 \\ \hline
    \end{tabular}
\end{table}

Three mixtures at different mass percent composition ratios (CRs) are constructed, as shown in Table~\ref{tab:systems} where ``Total'' refers to the total number of solvent beads in a system. All the mixtures are placed in a rectangular cell of dimensions $L_x \times L_y \times L_z$ to establish a liquid-vapor coexistence state with an interface parallel with the $xy$-plane. For the results reported here, $L_x = L_y = 88\sigma$ and $L_z = 287\sigma$. An example of an equilibrium state is shown in Fig.~\ref{fig:box}(a) for the mixture CR$_{1}$, which has an equal mass fraction of monomers and dimers. Periodic boundary conditions are imposed in the $xy$ plane. Along the $z$-direction, the mixture is confined between two flat walls, one below the liquid phase at $z=0$ and one above the vapor phase at $z=L_z$. The solvent-wall interaction is governed by the following LJ 9-3 potential,
\begin{equation} \label{eqn:wall}
    U_w(r_\perp)=\epsilon_w \left[\frac{2}{15}\left(\frac{\sigma_w}{r_\perp}\right)^{9}
    -\left(\frac{\sigma_w}{r_\perp}\right)^{3}\right]~,
\end{equation}
where $r_\perp$ is the perpendicular distance of a solvent bead from the wall. For the solvent-wall interaction, $\epsilon_w = 1.0\epsilon$ and $\sigma_w = 1.0\sigma$. At the lower wall, the potential is truncated at $3.0\sigma$ so that the liquid mixture adheres to the wall. The interaction between the solvent beads and the upper wall is purely repulsive with a cutoff distance of $r_c=0.858\sigma$. At equilibrium, the liquid-vapor interface is $\sim 115\sigma$ away from the top wall and the liquid film has a thickness of $\sim 170\sigma$.


To implement evaporation, a slab within $20\sigma$ below the top wall is designated as the deletion zone. To model evaporation into a vacuum, all beads in the deletion zone are removed every $\tau$. To model slower evaporation at a fixed rate, a small number (4 or 12) of solvent beads are removed every $\tau$. The evaporation rate, $j_e(t)$, is defined as the number of beads removed per unit area and time, which can be computed as
\begin{equation}
    j_e(t) = \frac{1}{L_x L_y}\frac{\text{d}N_e(t)}{\text{d}t}~,
\end{equation}
where $N_e(t)$ is the number of the removed beads accumulated over time.

All simulations are performed with the Large-scale Atomic/Molecular Massively Parallel Simulator (LAMMPS).\cite{LAMMPS} The equations of motion are integrated with a velocity-Verlet algorithm at a time step of $\delta t=0.01\tau$. All the solvent beads in the simulation box are thermalized at a fixed temperature ($T_b$) using a dissipative particle dynamics (DPD) thermostat, which conserves momentum locally. The DPD thermostat is implemented through a pairwise dissipative force. For a pair of solvent beads, $i$ and $j$, at a separation $r_{ij}$ within a cutoff distance $r_c$, the magnitude of the dissipative force is given by \cite{groot1997}
\begin{eqnarray}
    F(r_{ij}) &=& -\gamma \left(1-\frac{r_{ij}}{r_{c}}\right)^{2}\left(\hat{r}_{ij}\cdot\vec{v}_{ij}\right)+ \nonumber \\
    & &  +\mu \chi \left(1-\frac{r_{ij}}{r_{c}}\right)\left(\delta t\right)^{-\frac{1}{2}}~,
\end{eqnarray}
where $\gamma$ is a friction coefficient, $\hat{r}_{ij}$ is a unit vector pointing from bead $i$ to bead $j$, $\vec{v}_{ij}$ is the relative velocity of bead $i$ with respect to bead $j$, $\mu = \sqrt{2k_\text{B}T_b \gamma}$ with $k_\text{B}$ being the Boltzmann constant, and $\chi$ is a Gaussian random number with zero mean and unit variance. A force with the above magnitude and in the direction of $\hat{r}_{ij}$ is applied to bead $j$ while an opposite force is applied to bead $i$. The sum of the dissipative forces between any pair of solvent beads is therefore 0 and the total momentum of the pair is not influenced by the thermostat. This feature guarantees that the local evaporation behavior of the solvent is properly captured.

The simulation results using the DPD thermostat are further compared with those in which a weak Langevin thermostat is applied only to those solvent beads in a thin slab ($\sim 20 \sigma$) adjacent to the bottom wall.\cite{Cheng2011, Tang2018Langmuir, Tang2019Langmuir_control} The comparison corroborates the validity of utilizing a DPD thermostat to control the bulk temperature of an evaporating liquid (see Fig.~\ref{fig:thermo} below).

After the binary solvent CR$_{1}$ is equilibrated at $T_b =1.0\epsilon/k_\text{B}$, a binary mixture of nanoparticles with radius $a = 2.5\sigma$ is added to the liquid phase of the solvent. The initial volume fraction is about $4.5\%$ for each type of nanoparticles. The nanoparticles interact with each other via an integrated LJ potential between two spheres, which can be decomposed into an attractive and a repulsive component \cite{Everaers2003},
\begin{equation}\label{eqn:col_col_total}
    U_\text{CC}(r)=U_\text{A}(r)+U_\text{R}(r)~,
\end{equation}
where $r$ is the center-to-center separation of two nanoparticles. The functional forms of the attractive term is \cite{Everaers2003} 
\begin{eqnarray}\label{eqn:col_col_attr} 
    U_\text{A}(r) &=& -\frac{A_\text{CC}}{6}\left[\frac{2a_{1}a_{2}}{r^{2}-(a_{1}+a_{2})^{2}} +\frac{2a_{1}a_{2}}{r^{2}-(a_{1}-a_{2})^{2}} \right. \nonumber \\
    & & \left. 
    +\ln\left(\frac{r^{2}-(a_{1}+a_{2})^{2}}{r^{2}-(a_{1}-a_{2})^{2}}\right)\right]~,
\end{eqnarray}
where $A_{CC}$ is a Hamaker constant controlling the strength of interaction, and $a_1$ and $a_2$ are the radii of the two nanoparticles involved in the interaction, respectively. In this work, $A_\text{CC} = 39.5\epsilon$, and $a_1 = a_2 = a =2.5\sigma$. The repulsive component of the nanoparticle-nanoparticle interaction is
\begin{eqnarray}\label{eqn:col_col_rep}
    U_\text{R}(r) &=& \frac{A_\text{CC}}{37800}\frac{\sigma^{6}}{r} \nonumber \\
    & & \times \left[\frac{r^{2}-7r(a_{1}+a_{2})+6(a_{1}^{2}+7a_{1}a_{2}+a_{2}^{2})}{(r-a_{1}-a_{2})^{7}} \right. \nonumber \\
    & & +\frac{r^{2}+7r(a_{1}+a_{2})+6(a_{1}^{2}+7a_{1}a_{2}+a_{2}^{2})}{(r+a_{1}+a_{2})^{7}} \nonumber \\
    & & -\frac{r^{2}+7r(a_{1}-a_{2})+6(a_{1}^{2}-7a_{1}a_{2}+a_{2}^{2})}{(r+a_{1}-a_{2})^{7}} \nonumber \\
    & & \left. -\frac{r^{2}-7r(a_{1}-a_{2})+6(a_{1}^{2}-7a_{1}a_{2}+a_{2}^{2})}{(r-a_{1}+a_{2})^{7}}\right]~.
\end{eqnarray}
The potential $U_\text{CC}(r)$ is truncated at $r_c = 5.595\sigma$ so that the interaction between nanoparticles is purely repulsive and the nanoparticles can be easily dispersed in the binary solvent to form a uniform solution.

The interaction between a solvent bead and a nanoparticle is given by an additional integrated LJ potential between a point mass and a sphere,\cite{Everaers2003}
\begin{eqnarray}\label{eqn:col_sol}
    U_\text{CS}(r) & = & \frac{2A_\text{CS}}{9} \frac{a^{3}\sigma^{3}}{(a^{2}-r^{2})^3} \times \left[1- \right. \nonumber \\
    & & \left. \frac{(5a^{6}+45a^{4}r^{2}+63a^{2}r^{4}+15r^{6})\sigma^{6}}{15(a-r)^{6}(a+r)^6}\right]~.
\end{eqnarray}
Here $A_\text{CS}$ is another Hamaker constant. The two types of nanoparticles, labeled $\alpha$ and $\beta$, are identical except for their interaction strengths with the two solvent components, which are controlled by the corresponding Hamaker constants. In this work, for the $\alpha$-monomer and $\beta$-dimer interactions, $A_{\alpha m} = A_{\beta d} =100\epsilon$ while for the $\alpha$-dimer and $\beta$-monomer interactions, $A_{\alpha d} = A_{\beta m} =150\epsilon$. That is, the $\alpha$-type nanoparticles interact more strongly with the dimers while the $\beta$-type nanoparticles have stronger interactions with the monomers. Since all the nanoparticles have a radius of $2.5\sigma$, the solvent-nanoparticle interaction is truncated at $r_c=6.5\sigma$ to include the attractive tail of $U_\text{CS}(r)$ so that the nanoparticles are well solvated by the binary solvent.

All nanoparticles are confined in the simulation box by the same two walls used to confine the solvent beads. The nanoparticle-wall interaction is also governed by the LJ 9-3 potential in Eq.~(\ref{eqn:wall}) with $\epsilon_w = 2.0 \epsilon$ and $\sigma_w = a = 2.5\sigma$. The potential is truncated at $r_c=0.858\sigma_w$ at both top and bottom walls so that the nanoparticles do not adhere to the walls.


With the DPD thermostat only applied to the solvent beads, the system is equilibrated for at least $2\times 10^5 \tau$ prior to evaporation. A visualization of one equilibrated solution is shown in Fig.~\ref{fig:box}(b). Evaporation is initiated by removing solvent beads in the deletion zone as described earlier. Three schemes are studied. In the first, all solvent beads in the deletion zone are removed every $\tau$, mimicking evaporation into a vacuum where the evaporation rate is initially high, then decreases over time, and eventually reaches a plateau value of $j_p$. In the second, $12$ solvent beads in the deletion zone are randomly removed every $\tau$ to yield a constant evaporation rate of $j_p$. This is referred to as the intermediate-rate scheme. In the slow-rate scheme, 4 solvent beads are removed from the deletion zone every $\tau$, corresponding to an evaporation rate $j_s \simeq j_p/3$. Since the entire dimer is removed if a dimer bead is randomly selected for deletion, the actual number of removed beads every $\tau$ can sometimes be slightly more than 4 in the slow-rate scheme or 12 in the intermediate-rate scheme.

\section{Results and Discussion}\label{sec:res}

\subsection{Evaporation of Monomer-Dimer Mixtures}

\begin{figure}[htb]
    \centering
    \includegraphics[width=\columnwidth]{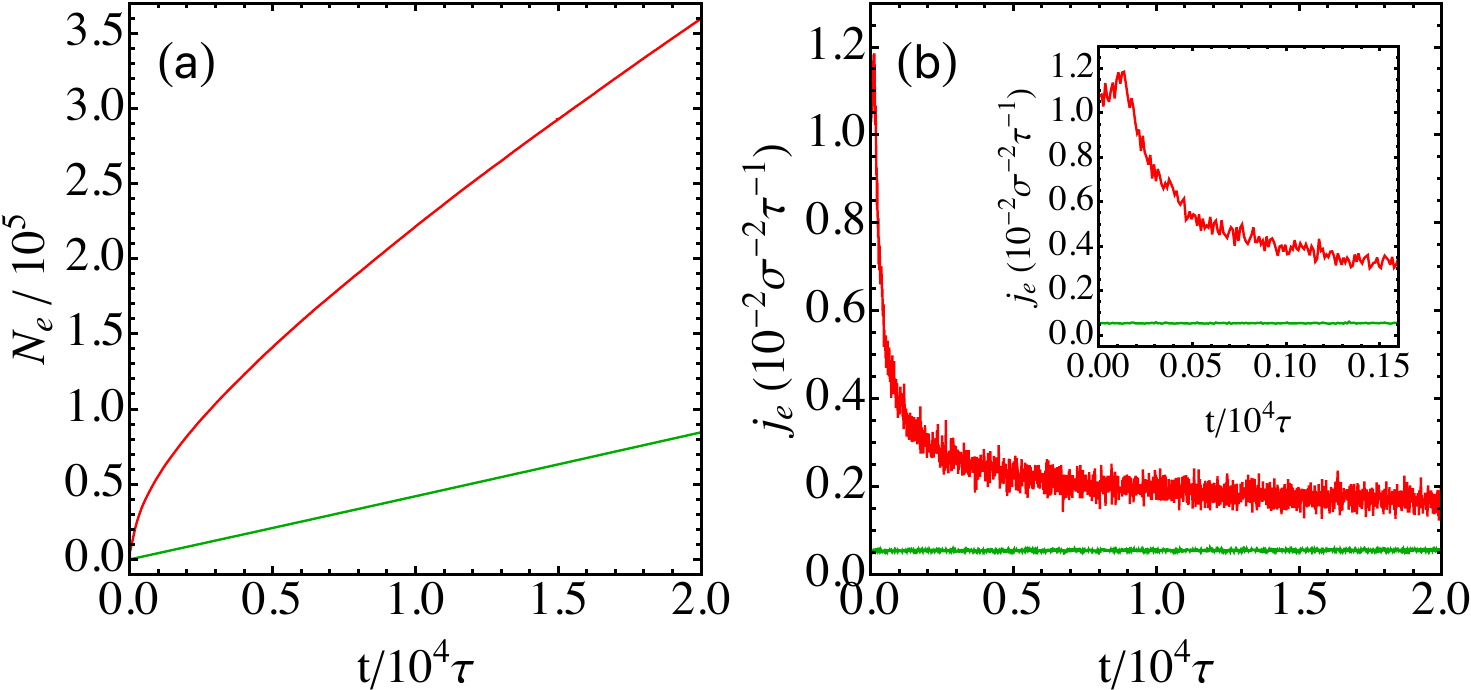}
    \caption{(a) Accumulated number of evaporated beads ($N_e$) and (b) evaporation rate ($j_e$) vs. time ($t$). Data are for CR$_1$ at $T_b = 1.0\epsilon/k_\text{B}$ evaporating into a vacuum (red) or at a fixed rate (green, $j_e \simeq 5.5\times 10^{-4}\sigma^{-2}\tau^{-1}$). The inset of (b) shows the variation of $j_e$ over time from the start of evaporation to $1.5\times 10^3\tau$.}
    \label{fig:evap_rate}
\end{figure}

The mixture CR$_1$ is used first to illustrate the behavior of mixed binary solvents under the fastest and slowest evaporation schemes. The accumulated number of evaporated beads, $N_e(t)$, is plotted as a function of time in Fig.~\ref{fig:evap_rate}(a), where the time when evaporation is initiated is designated as $t=0$. The corresponding evaporation rate, $j_e(t)$, is plotted in Fig.~\ref{fig:evap_rate}(b). Note that $j_e(t)$ is the total evaporation rate of both monomers and dimers. 

In the evaporating-into-vacuum scheme, the evaporation rate is initially large ($j_e \simeq 1.0 \times 10^{-2} \sigma^{-2}\tau^{-1}$) and even increases slightly to about $j_e \simeq 1.2 \times 10^{-2} \sigma^{-2}\tau^{-1}$ since the saturation vapor density of LJ monomers is high. A similar behavior was observed previously for an evaporating liquid consisting purely of LJ monomers.\cite{Cheng2011} As evaporation proceeds further, the evaporation rate of the monomer-dimer mixture decreases with time as the vapor gets depleted quickly and the rate eventually flattens into a plateau value of $j_p \simeq 1.7 \times 10^{-3} \sigma^{-2}\tau^{-1}$, which is a factor of 7 smaller than the peak value. In the slow-rate scheme, $N_e(t)$ essentially grows linearly with time and the resulting evaporation rate has a constant value of $j_s \simeq 5.5 \times 10^{-4} \sigma^{-2}\tau^{-1}$, which is as expected slightly higher than the rate ($\sim 5.2 \times 10^{-4} \sigma^{-2}\tau^{-1}$) corresponding to removing exactly 4 beads every $\tau$. The slow rate $j_s$ is about $1/3$ of $j_p$, the plateau rate in the evaporating-into-vacuum scheme.

\begin{figure*}[htb]
    \includegraphics[width=\textwidth]{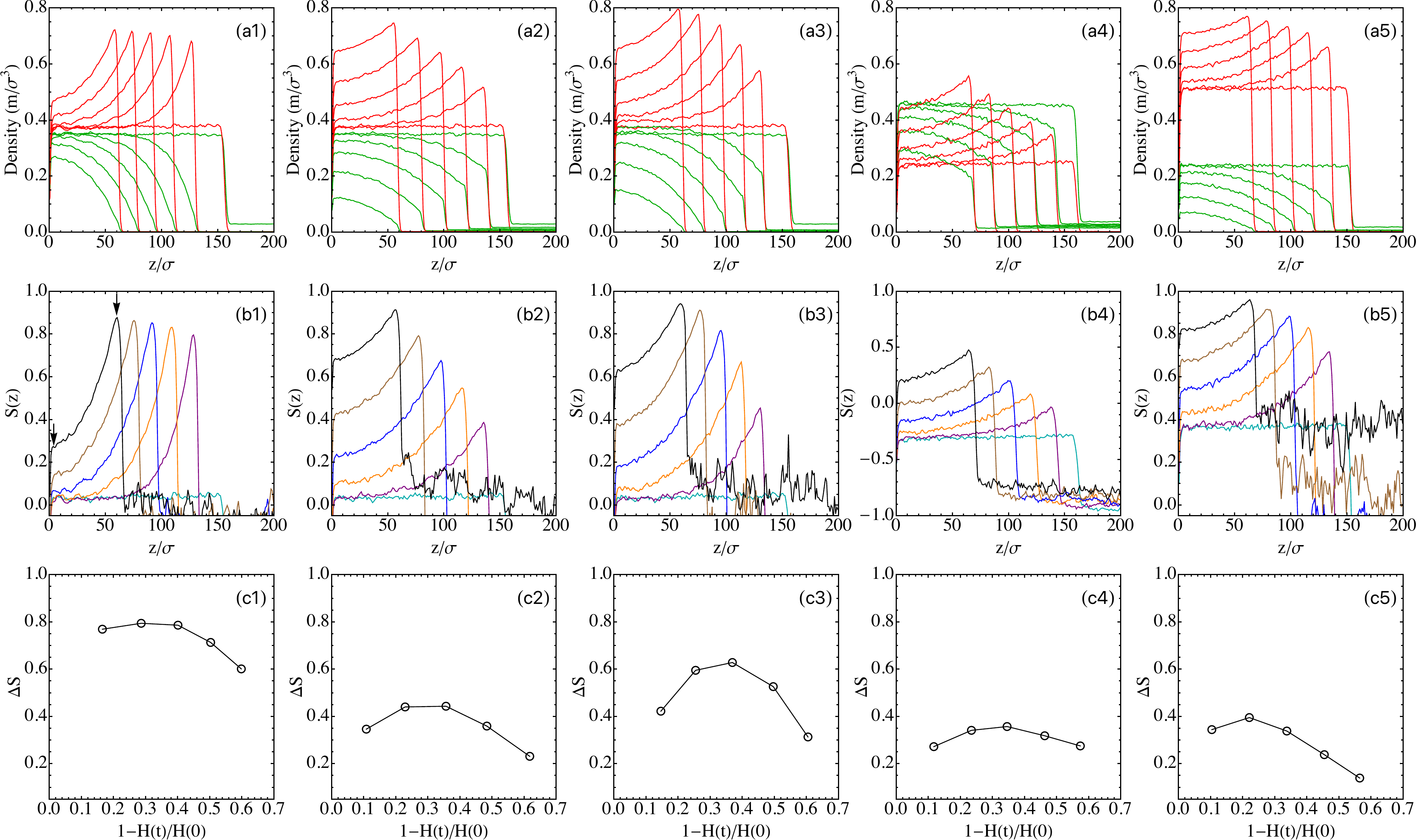}
    \caption{Top row: the density profiles of dimer beads (red) and monomer beads (green) as a function of $z$ at various times for the 5 evaporating systems. Middle row: the corresponding local order parameter of inhomogeneity between the two solvent components. Bottom row: the overall magnitude of inhomogeneity as a function of the extent of drying. From the first column on the left to the fifth column on the right, the 5 evaporating systems are: CR$_1$ at $T_b = 1.0\epsilon/k_\text{B}$ evaporating into a vacuum; CR$_1$ at $T_b = 1.0\epsilon/k_\text{B}$ evaporating at the slow rate $j_s$; CR$_1$ at $T_b = 0.9\epsilon/k_\text{B}$ evaporating at $j_s$; CR$_2$ at $T_b = 1.0\epsilon/k_\text{B}$ evaporating at $j_s$; CR$_{0.5}$ at $T_b = 1.0\epsilon/k_\text{B}$ evaporating at $j_s$.}
    \label{fig:den_op_mag}
\end{figure*}

Although LJ dimers and monomers are miscible, they differ in their vapor density and volatility.\cite{Cheng2011} As a result, their distribution in the drying mixture may become nonuniform. To understand the behavior of monomers and dimers during evaporation, 5 systems are studied with various composition ratios and bulk temperatures under either the evaporating-into-vacuum scheme or the slow-rate scheme. The density profiles of dimers and monomers at 6 evenly spaced times, starting at $t=0$ when evaporation is initiated, are plotted in the top row of Fig.~\ref{fig:den_op_mag}. With a lower volatility, dimers start to concentrate near the evaporation front, leading to a positive density gradient in the liquid domain. Monomers have a larger saturation vapor density and evaporate faster than the dimers. Consequently, monomers become depleted and their density decreases as the liquid-vapor interface is approached from the liquid side.

As shown in Fig.~\ref{fig:den_op_mag}, the distribution of dimers and monomers in the mixture develops inhomogeneity during evaporation. To characterize this inhomogeneity, a local order parameter $S(z)$ is introduced,
\begin{equation}
    S(z) = \frac{\rho_d(z)-\rho_m(z)}{\rho_d(z)+\rho_m(z)}~,
\end{equation}
where $\rho_d(z)$ and $\rho_m(z)$ are the local number density of dimer beads and monomer beads, respectively, in a slab in the $xy$-plane spanning from $z-\Delta z$ to $z+\Delta z$ with $\Delta z = 0.5\sigma$. In a uniform mixture, $S(z)$ is a constant. A positive $S(z)$ indicates that in the local region there are more dimer beads than monomer beads while a negative $S(z)$ corresponds to monomer beads having a higher local concentration than dimer beads. The results on $S(z)$ for the 5 evaporating systems are plotted in the second row of Fig.~\ref{fig:den_op_mag}. As expected, in the equilibrium mixtures prior to evaporation, $S(z)$ has a constant value across the liquid domain. During evaporation, $S(z)$ varies with $z$ and its local value indicates the extent of nonuniformity of the local distribution of dimers and monomers. In all cases, $S(z)$ increases with $z$ in the liquid phase and reaches a peak value at the receding liquid-vapor interface, where the difference between $\rho_d(z)$ and $\rho_m(z)$ is most significant.

To further quantify the extent of inhomogeneity developed in the dimer-monomer distribution during evaporation, the difference between the peak value of $S(z)$ at the liquid-vapor interface and the value of $S(z)$ for the mixture adjacent to the bottom confining wall at $z=0$ is taken as the overall magnitude of inhomogeneity, $\Delta S$. One example is shown in Fig.~\ref{fig:den_op_mag}(b1), where the peak and base values are taken at the locations indicated by the two arrows. The difference of the two values yields $\Delta S$.

The results for $\Delta S$ as a function of the extent of drying are plotted in the third row of Fig.~\ref{fig:den_op_mag}. The extent of drying is defined as $1-H(t)/H(0)$, where $H(t)$ is the thickness of the liquid domain along the $z$-axis at an elapse time of $t$ after the initiation of evaporation and $H(0)$ is the equilibrium thickness. At $t=0$, $1-H(t)/H(0)=0$ and the overall magnitude of inhomogeneity ($\Delta S$) is obviously 0 by definition. As shown in Fig.~\ref{fig:den_op_mag}, $\Delta S$ exhibits a similar trend with time for all 5 evaporating systems. It first grows quickly as evaporation starts, then reaches a maximum value ($\Delta S_\text{max}$) at $1-H(t)/H(0) \simeq 0.3$ (i.e., at $H(t)\simeq 0.7 H(0)$), and finally decreases with time in the late stage of drying. The maximum value of $\Delta S$ varies significantly from one system to another. For the CR$_1$ mixture at $T_b = 1.0\epsilon/k_\text{B}$ evaporating into a vacuum, $\Delta S_\text{max} \simeq 0.8$. For the same CR$_1$ mixture at $T_b = 1.0\epsilon/k_\text{B}$ evaporating at a fixed small rate of $j_s$, $\Delta S_\text{max}$ is only around $0.44$. The comparison of these two cases indicates that the inhomogeneity of the dimer-monomer distribution during evaporation is larger at faster evaporation. For the CR$_1$ mixture evaporating at $j_s$ but with $T_b$ lowered to $0.9\epsilon/k_\text{B}$, the variation of $\Delta S$ with time is more pronounced and $\Delta S_\text{max} \simeq 0.63$. This can be understood by noticing that the same evaporation rate $j_s$ is a relatively faster rate for the system at $T_b = 0.9\epsilon/k_\text{B}$ than for the same system at $T_b = 1.0\epsilon/k_\text{B}$, as the former has a lower vapor density.

The results in the last two columns of Fig.~\ref{fig:den_op_mag} are for the CR$_{0.5}$ and CR$_2$ mixtures, respectively. Both systems are evaporated at a fixed rate of $j_s$ with $T_b = 1.0\epsilon/k_\text{B}$. The value of $\Delta S_\text{max}$ is $\sim 0.36$ for the former and $0.4$ for the latter, slightly smaller than the corresponding value for the CR$_1$ mixture under the same evaporation condition. Furthermore, for the CR$_2$ mixture, $\Delta S$ decreases with the extent of drying more quickly after reaching its maximum value, which is caused by the faster depletion of monomers as there are relatively fewer monomers in the CR$_2$ mixture.

The results presented so far are all obtained using a DPD thermostat to maintain a constant temperature throughout the mixture. The DPD thermostat is adopted as it preserves momentum locally, which is important for a proper description of an evaporation process. However, it removes evaporative cooling effects. Another strategy, which was used in several previous studies,\cite{Cheng2011, Tang2018Langmuir, Tang2019Langmuir_control} is to only thermalize a thin layer of liquid adjacent to the bottom wall with a Langevin thermostat, mimicking a thermal contact with a substrate held at a constant temperature. That is, the solvent beads in the thin slab bounded below by the bottom wall move according to the Langevin equation while the solvent beads in the other region still follow Newton's equation of motion. In this approach, evaporative cooling naturally occurs and the temperature at the evaporating interface becomes lower than the ``bulk'' temperature in the thermalized layer, especially at high evaporation rates.

\begin{figure}[htb]
    \centering
    \includegraphics[width=\columnwidth]{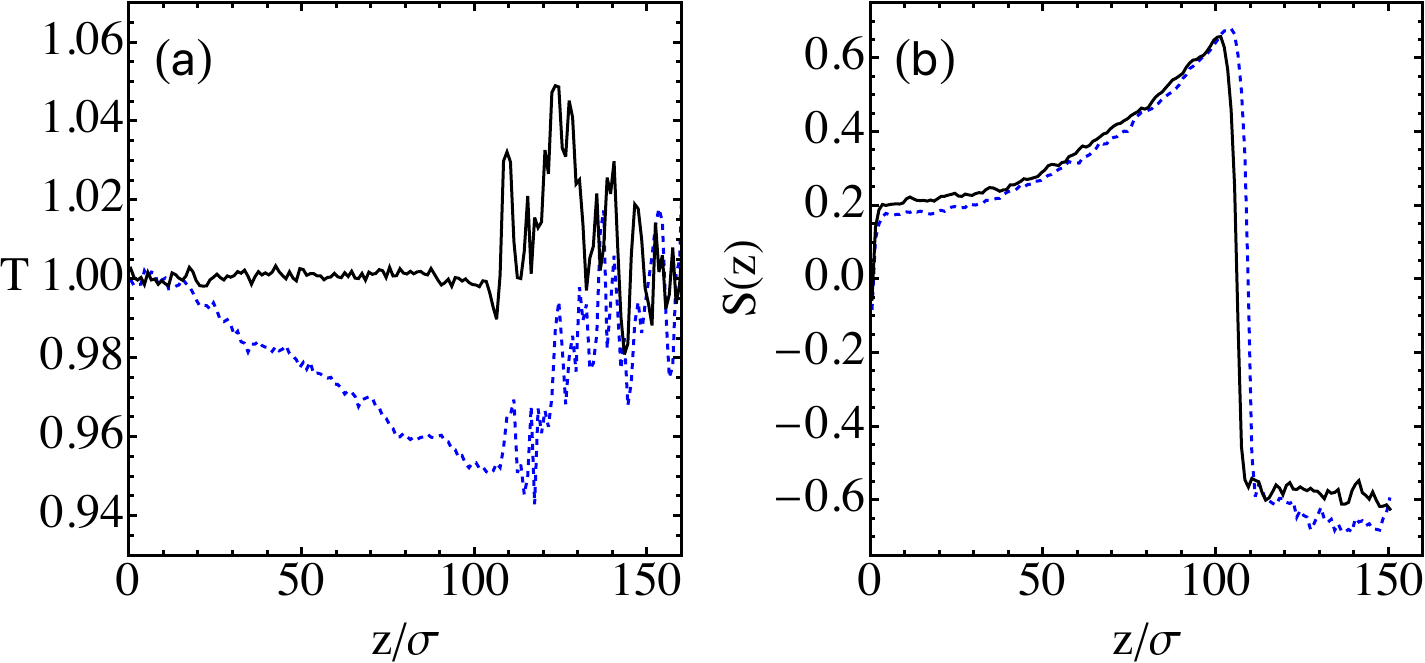}
    \caption{Comparison of (a) the temperature field and (b) the local order parameter of inhomogeneity as functions of $z$ from simulations based on the DPD thermostat (black solid line) and the Langevin thermostat (blue dashed line) when $\sim 60\%$ of the liquid solvent remains. The results are for the CR$_1$ mixture with $T_b = 1.0\epsilon/k_\text{B}$ evaporating at $j_s$.}
 \label{fig:thermo}
\end{figure}

In Fig.~\ref{fig:thermo}, two sets of results on the temperature field and the local order parameter of inhomogeneity, $S(z)$, along the $z$-axis are shown, with one set from the MD simulations based on the DPD thermostat while the other from the simulations in which the solvent beads within $20\sigma$ from the bottom wall are thermalized with a Langevin thermostat. The damping time in the Langevin thermostat is set to $10\tau$. The data are for the CR$_1$ mixture evaporating at $j_s$ with $T_b = 1.0\epsilon/k_\text{B}$. Here $T_b$ is the target temperature in the entire simulation box in the case with the DPD thermostat and in the solvent layer being thermalized in the case with the Langevin thermostat. As expected, evaporative cooling occurs in the latter case with the interfacial temperature lower than $T_b$ by about $5\%$ (see Fig.~\ref{fig:thermo}(a)). However, the profiles of $S(z)$ from the two simulations using the different thermostats essentially trace each other, as shown in Fig.~\ref{fig:thermo}(b). The minor visible difference between the 2 curves in Fig.~\ref{fig:thermo}(b) is caused by the slight difference in the extent of drying at which the data are collected. Such comparison corroborates the robustness of the results reported above on the evaporation behavior of dimer-monomer mixtures.

\subsection{Evaporation of Colloidal Suspensions with a Monomer-Dimer Mixture as Solvent}

The inhomogeneity in the monomer-dimer distribution developed during evaporation can be exploited to stratify nanoparticles in a drying film. To demonstrate this idea, we employ the CR$_1$ mixture at $T_b = 1.0\epsilon/k_\text{B}$ as a solvent. Two types of nanoparticles, labeled $\alpha$ and $\beta$, are dispersed in the liquid mixture and the resulting suspension is then equilibrated for $2\times 10^5\tau$. The number of each type of nanoparticles is $900$, yielding a volume fraction of $\sim 4.5\%$ for each nanoparticle species in the initial suspension. One snapshot of the equilibrated suspension is shown in Fig.~\ref{fig:box}(b), where the nanoparticles are clearly uniformly dispersed in the mixed solvent.

\begin{figure}[htb]
    \centering
    \includegraphics[width=\columnwidth]{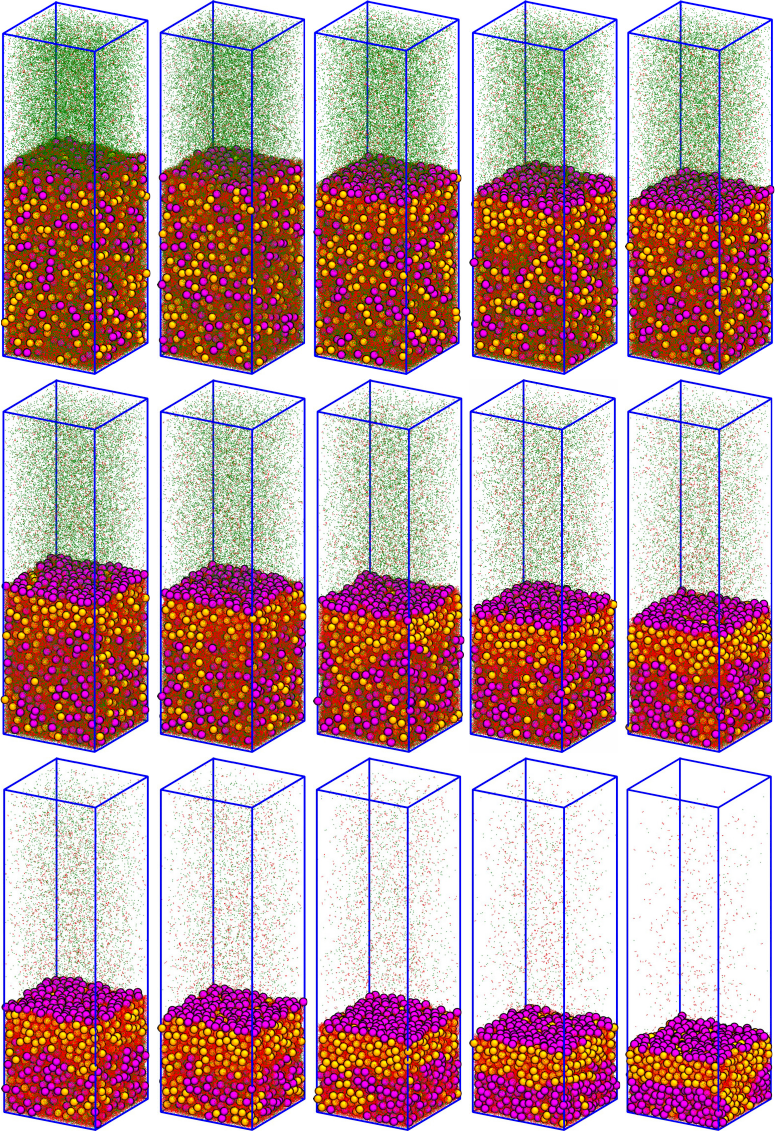}
    \caption{Snapshots showing the evolution of nanoparticle distribution ($\alpha$-type in orange and $\beta$-type in purple) in the drying film with the CR$_1$ mixture solvent evaporating at the rate of $j_s$. The snapshots are ordered from left to right and top to bottom with the first snapshot taken at $t=0$ when evaporation is initiated. The elapsed time between two consecutive snapshots is $1.9\times 10^4\tau$.}
    \label{fig:snapshot_evolution}
\end{figure}

Three evaporation schemes for the mixture solvent are investigated, including evaporating into a vacuum, evaporating at a fixed intermediate rate equal to the plateau rate $j_p$ in Fig.~\ref{fig:evap_rate}(b), and evaporating at an even slower rate of $j_s~(\simeq j_p/3)$. A series of snapshots showing the evolution of nanoparticle distribution in the drying film are shown in Fig.~\ref{fig:snapshot_evolution} for the third scheme. The results for the other two schemes are qualitatively similar. Prior to evaporation, nanoparticles are uniformly dispersed in the liquid film, though there is a slightly higher concentration of $\beta$-type nanoparticles at the liquid-vapor interface. The underlying reason is that at the liquid-vapor interface the concentration of monomers is higher than that of dimers as the monomers have a lower surface tension. Since the $\beta$-type nanoparticles interact more favorably with monomers, there are more $\beta$ nanoparticles at the equilibrium liquid-vapor interface than $\alpha$ nanoparticles.

As evaporation progresses, the enrichment of $\beta$ nanoparticles at the receding interface becomes more pronounced, as shown in the top row of Fig.~\ref{fig:snapshot_evolution}. At the same time, since dimers become concentrated while monomers get depleted near the interface during evaporation (see the top row of Fig.~\ref{fig:den_op_mag}), the $\alpha$-type nanoparticles start to accumulate just below the surface layer of $\beta$ nanoparticles that are trapped at the interface. Driven by the density gradient of monomers, other $\beta$-type nanoparticles initially dispersed in the bulk suspension start to move towards the lower part of the suspension. The final distribution of nanoparticles exhibit a stratified $\beta$-$\alpha$-$\beta$ sandwich structure. But it should be pointed out that stratification is not $100\%$ in the drying film as in the $\beta$-layer adjacent to the bottom wall, there are still domains of $\alpha$ nanoparticles. Sandwich structures have also been observed in the experiment of Liu \textit{et al.} on bidisperse nanoparticle suspension films being dried quickly,\cite{LiuWeiping2019} where depending on the size ratio of the binary nanoparticles, large-small-large or small-large-small sandwich layering were found.

\begin{figure}[htb]
    \centering
    \includegraphics[width=\columnwidth]{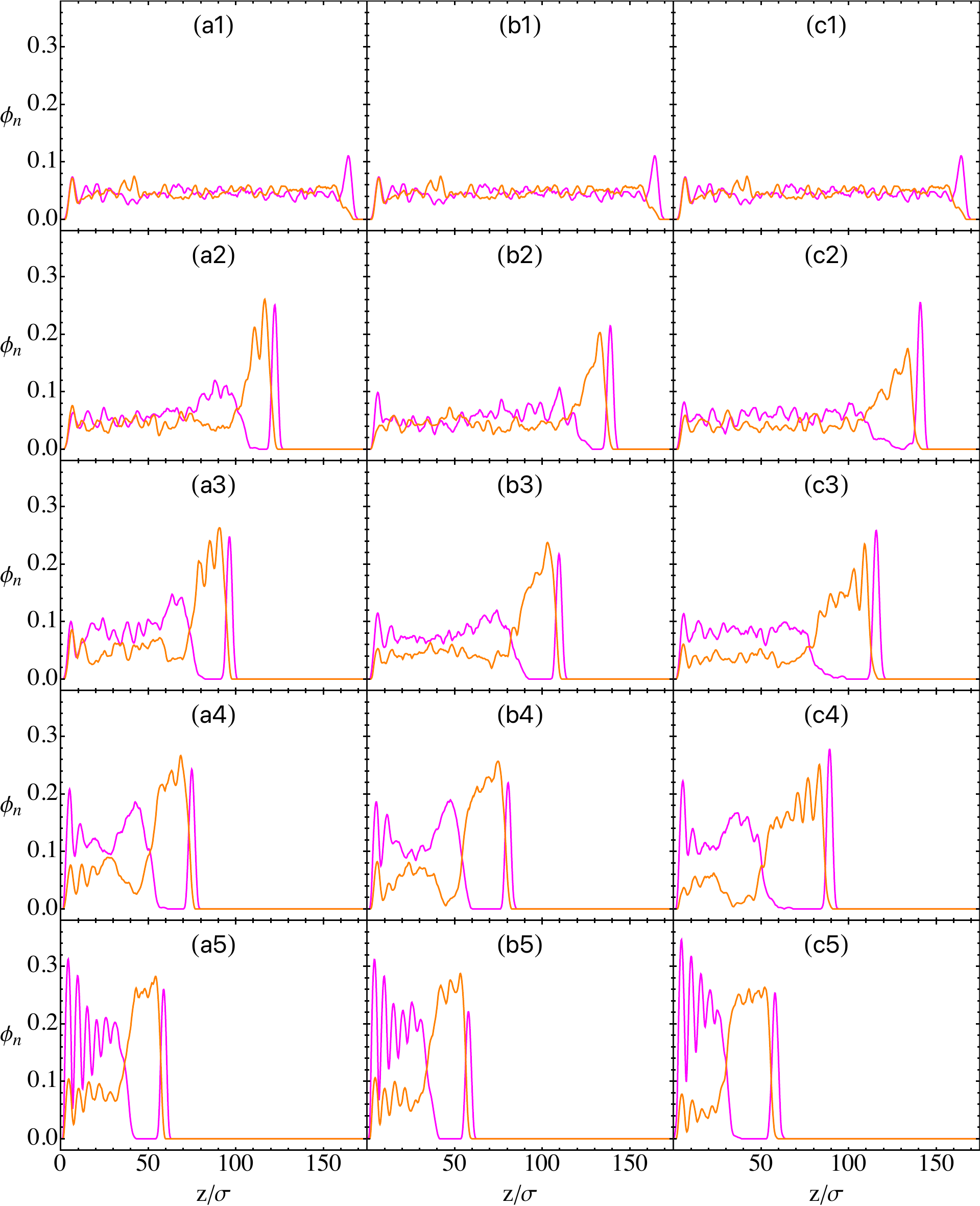}
    \caption{Volume concentration profiles of the two types of nanoparticles ($\alpha$ in orange and $\beta$ in purple) in the drying film under three solvent evaporation schemes: (a1)-(a5) evaporating into a vacuum, (b1)-(b5) evaporating at a fixed intermediate rate of $j_p$ (the terminal plateau rate in the evaporating-into-vacuum scheme), (c1)-(c5) evaporating at a fixed rate of $j_s \simeq j_p/3$. All three runs start with the same initial equilibrium state at $t=0$ (first row). Panels (a2)-(a5) are for $t=2.1\times 10^4\tau$, $4.2\times 10^4\tau$, $6.3\times 10^4\tau$, and $8.4\times 10^4\tau$; (b2)-(b5) are for $t=0.3\times 10^5\tau$, $0.6\times 10^5\tau$, $0.9\times 10^5\tau$, and $1.2\times 10^5\tau$; (c2)-(c5) are for $t=0.65\times 10^5\tau$, $1.3\times 10^5\tau$, $1.95\times 10^5\tau$, and $2.6\times 10^5\tau$, respectively. In all cases, the final film thickness is about $62\sigma$, as shown in the last row.}
    \label{fig:den_np}
\end{figure}

The evolution of nanoparticle distribution in the drying film is quantified as concentration profiles along the $z$-axis shown in Fig.~\ref{fig:den_np}. Here a volume concentration ($\phi_n$) is used. Considering a slab in the $xy$-plane spanning from $z-\Delta z$ to $z-\Delta z$, $\phi_n(z)$ refers to the fraction of slab volume occupied by each type of nanoparticles. To compute $\phi_n(z)$, the volume of each nanoparticle needs to be partitioned to a series of adjacent slabs that the nanoparticle intersect with. Results on $\phi_n(z)$ at $\Delta z = 0.125\sigma$ for both $\alpha$ and $\beta$ nanoparticles under the three evaporation schemes are included in Fig.~\ref{fig:den_np}: evaporating-into-vacuum [Figs.~\ref{fig:den_np}(a1)-(a5)], fixed intermediate rate [Figs.~\ref{fig:den_np}(b1)-(b5)], and fixed slow rate [Figs.~\ref{fig:den_np}(c1)-(c5)].

All three evaporation simulations start with the same equilibrium suspension and the corresponding concentration profiles of nanoparticles are repeatedly shown in the first row of Fig.~\ref{fig:den_np}. In equilibrium, there is a clear concentration peak for the $\beta$-type nanoparticles at the liquid-vapor interface as the interface of $\beta$-favoring monomers protrudes that of $\alpha$-favoring dimers. The subsequent evolution of concentration profiles shown in Fig.~\ref{fig:den_np} reveals interesting trends as the solvent evaporation is slowed down. In all cases, the $\beta$ concentration peak at the evaporation front grows quickly and then saturates after the solvent evaporation is turned on. Right below this surface $\beta$-layer, $\alpha$ nanoparticles begin to concentrate as evaporation proceeds and the degree of concentration is stronger in the evaporating-into-vacuum scheme, which has the fastest solvent evaporation rate. At the same time, $\alpha$ nanoparticles are depleted and $\beta$ nanoparticles are enriched in the lower part of the film. In the evaporating-into-vacuum scheme, a second smaller concentration peak of $\beta$ nanoparticles emerge below the concentrated region of $\alpha$ nanoparticles [see Fig.~\ref{fig:den_np}(a2)]. A similar peak, though even weaker, appears at a progressively later time under slower evaporation [see Figs.~\ref{fig:den_np}(b3) and (c4)]. This trend is an outcome of the weakening density inhomogeneity of the two solvent components and thus a weaker drive to move nanoparticles when the evaporation rate is reduced.

The concentration profiles in Fig.~\ref{fig:den_np} confirm that the two types of nanoparticles eventually form the $\beta$-$\alpha$-$\beta$ sandwich distribution, which is easily detected in the film of a thickness of $\sim 62\sigma$ (see the last row of Fig.~\ref{fig:den_np}). In the middle $\alpha$-layer, there are no $\beta$ nanoparticles but in the lower $\beta$-dominated layer, $\alpha$ nanoparticles are still present. The nanoparticle concentration also exhibits oscillation near the bottom wall, reflecting the epitaxial order induced by the wall.\cite{Thompson1990}

The trends reflected in Fig.~\ref{fig:den_np} can be further understood by examining the competition between the convective motion of nanoparticles imposed by solvent evaporation and their diffusion. The former is characterized by the receding speed, $v_e$, of the liquid-vapor interface and the latter by the nanoparticle diffusion coefficient, $D$. Independent simulations were performed to compute $D$ for both $\alpha$ and $\beta$ nanoparticles in the CR$_1$ mixture with the same particle volume fractions as in the equilibrium suspension prior to evaporation. For the $\alpha$-type, $D_\alpha \simeq 4.5\times 10^{-3}\sigma^2/\tau$ and for the $\beta$-type, $D_\beta \simeq 4.8\times 10^{-3}\sigma^2/\tau$. The two are close and we will simply use $D = (D_\alpha + D_\beta)/2 \simeq 4.65\times 10^{-3}\sigma^2/\tau$ for the discussion here. The receding speed of the liquid-vapor interface is controlled by the solvent evaporation rate (see Supporting Information). Under the evaporating-into-vacuum scheme, $v_e \simeq 1.4\times 10^{-3}\sigma/\tau$; for the fixed intermediate rate, $v_e \simeq 0.92\times 10^{-3}\sigma/\tau$; for the fixed slow rate, $v_e \simeq 0.40\times 10^{-3}\sigma/\tau$.

The ratio $D/v_e$ sets a characteristic length scale, $l_e$, which represents the thickness of the zone at the receding interface that is significantly perturbed by solvent evaporation. For the three evaporation schemes, $l_e \simeq 3.3,~5.1$, and $11.5\sigma$, respectively. As expected, $l_e$ is larger at slower evaporation. The middle three rows of Fig.~\ref{fig:den_np} also reflects this trend. As the evaporation rate is reduced, the thickness of the middle $\alpha$-rich layer gets larger, where the two types of nanoparticles are driven into phase separation by the contrasting density gradients of monomers and dimers. In the later stage of evaporation, the evaporation rate in the evaporating-into-vacuum scheme approaches the intermediate rate, $j_p$. As a result, the thickness of the middle $\alpha$-layer becomes comparable in the two cases, which is about $21\sigma$ for the drying film of a thickness of $\sim 62\sigma$ (see Figs.~\ref{fig:den_np}(a5) and (b5) and Supporting Information). On the other hand, under the slow evaporation rate $j_s$ ($\simeq j_p/3$), the $\alpha$-layer has a thickness of about $26\sigma$ (see Fig.~\ref{fig:den_np}(c5) and Supporting Information). In this case, the volume concentration of $\alpha$ nanoparticles in the lower $\beta$-layer is also reduced.

Figure~\ref{fig:den_np} indicates that the degree of stratification is enhanced in the slow-rate evaporation scheme, though the overall magnitude of inhomogeneity of the monomer-dimer distribution under this scheme is much smaller than that in the evaporating-into-vacuum scheme (see Fig.~\ref{fig:den_op_mag}). When the solvent evaporates more slowly, the nanoparticles have more time to diffuse and phase separate more strongly according to the distribution of dimers and monomers, thus improving stratification. However, at very slow evaporation, dimers and monomers are expected to be always uniformly distributed in the evaporating mixture and the inhomogeneity of their distribution does not develop. In this case, there is no driving force for the two types of nanoparticles to phase separate as they are identical except for their coupling strengths with the two solvent components. Therefore, stratification of nanoparticles is not expected to occur at all at very slow evaporation rates. To summarize, there exists an optimal evaporation rate at which the degree of nanoparticle stratification is maximized. This echoes a previous finding that for size-driven stratification in drying colloidal suspensions, the degree of stratification reaches a maximum value at some optimized evaporation rate.\cite{Tang2019JCP_compare} The existence of an optimal rate in evaporation-induced ordering was also found in a previous simulation study of nanoparticle assembly driven by solvent evaporation.\cite{Cheng2013}

\section{Conclusions}\label{sec:conc}

Evaporation-driven auto-stratification is potentially a useful technique for fabricating layered materials by dispersing particles in a proper solvent and then drying the resulting suspension under appropriate conditions. Not surprisingly, approaches to induce and control stratification are of great interest. It was previously demonstrated that for a bidisperse particle suspension, stratification can be tuned from large-on-top to small-on-top by imposing a temperature gradient in the direction of drying and utilizing the accompanying thermophoretic response of the suspended particles.\cite{Tang2019Langmuir_control} In this paper, a different approach is discussed, where a binary solvent mixture is employed to drive a binary blend of nanoparticles of the same size to stratify through contrasting couplings between the two types nanoparticles and the two solvent components. 

In the equilibrium state, there is a relatively higher concentration of the more volatile solvent at the liquid-vapor interface, which results in a thin layer of nanoparticles, favored by the more volatile solvent, at the liquid/vapor interface. When evaporation starts, the top layer of the more volatile solvent quickly evaporates and the less volatile solvent becomes concentrated at the interface. As a result, the thin layer of nanoparticles which favors the more volatile solvent are trapped at the interface (i.e., the evaporation front) by the amassing of mainly the other type of nanoparticles in the zone just below the liquid-vapor interface. This inhomogeneity in the distribution of the two solvents thereby prompts stratification of a binary mixture of nanoparticles even if the size of the two nanoparticles are comparable. As evaporation proceeds, right below the thin layer of the trapped nanoparticles at the interface, the nanoparticles that interact more strongly with the less volatile solvent become enriched while the nanoparticles that interact more strongly with the more volatile solvent are driven towards the bottom of the drying film. The final distribution of the two types of nanoparticles becomes a sandwich film.

Using a solvent mixture is a frequently attempted technique for processing materials.\cite{Abdulla-Al-Mamun2009,Gordon2016,Guo2008PRL,Hoogenboom2008,Iyengar2016,Ye2012AM,Shi2019ACSAMI,OConnell2019} The results presented here shows that a mixed binary solvent can be used to induce nanoparticle to develop stratified layered distributions during evaporation. This approach is more facile and feasible than the previous suggestion based on thermophoresis\cite{Tang2019Langmuir_control} as it is hard to induce and maintain a desirable temperature gradient in an evaporating suspension. It is hoped that the reported simulations can motivate experimental work to realize the control of stratification with a mixed solvent. To this end, at least 3 factors need to be considered and fulfilled. First, two solvent components in the mixture must have sufficiently different volatilities so that they evaporate at very different rates. Second, the different types of nanoparticles must have strikingly different coupling strengths with the different solvent components. Experimentally, an asymmetric coupling between nanoparticles and solvent components can be achieved by coating the nanoparticles with different materials such as polymers. Third, overall solvent evaporation must be rapid enough so that the distribution of the different solvent components become inhomogeneous during evaporation. When these conditions are satisfied, it expected that the differing density gradients of the various solvent components developed during evaporation can be utilized to drive nanoparticles to form a drying film with a stratified distribution.

The stratification scheme proposed here relies on the condition that the mixed solvent is spatially uniform in the plane perpendicular to the evaporation direction. That is, evaporation needs to be essentially one-dimensional in the direction of the evaporating film's thickness and translationally invariant along its surface. Otherwise, various instabilities can emerge,\cite{Bassou2009,Boulogne2015SM,Mitov1998} which are often associated with composition gradients parallel to the film surface and thermal gradients. These gradients can lead to local variations of the liquid-vapor interfacial tension that drive Marangoni flows.\cite{Yiantsios2006} They can also drive B\'{e}nard-Marangoni convection.\cite{Mitov1998} Fractures and cracks can appear as well,\cite{Singh2007,Boulogne2015SM,Badar2022} When these instabilities occur. The structure of the resulting dry film will be more complicated than (one-dimensional) stratification. However, it would also offer opportunities for fabricating colloidal materials beyond layered films. Understanding the mechanisms of various instabilities in a drying colloidal film and their effects on the film structure is still an exciting research direction.\cite{Kooij2015}

In the simulations reported here, only uncharged spherical particles are considered. Aspherical colloids are of interest as depletion interactions between particles may depend on their shapes.\cite{Onsager1949,Eisenriegler2005} Research on understanding the effect of particle shape in film formation during rapid solvent evaporation will be reported in the future. Colloidal particles are often charged, which can help stabilize colloidal suspensions and prevent aggregation. Solutions also often contain salts. The direction of motion of a colloidal particle under the density gradients of the solvent and other solutes depends on their mutual interactions, to which electrostatic interactions may contribute significantly.\cite{Keh2016} Therefore, a neutral hard-sphere-like system may behave very differently from its counterpart where the particles are charged. Furthermore, tuning electrostatic interactions may be another experimental approach to realize the asymmetric particle-solvent couplings utilized here to drive particle stratification. Clearly, more research is needed on understanding the drying behavior of colloidal suspensions of charged particles.

\subsection*{ACKNOWLEDGMENTS}

This material is based upon work supported by the National Science Foundation under Grant No. DMR-1944887. This research was initially supported by a 4-VA Collaborative Research Grant (``Material Fabrication via Droplet Drying''). This work was performed, in part, at the Center for Integrated Nanotechnologies, an Office of Science User Facility operated for the U.S. Department of Energy Office of Science. Sandia National Laboratories is a multimission laboratory managed and operated by National Technology and Engineering Solutions of Sandia, LLC., a wholly owned subsidiary of Honeywell International, Inc., for the U.S. Department of Energy's National Nuclear Security Administration under contract DE-NA0003525. This paper describes objective technical results and analysis. Any subjective views or opinions that might be expressed in the paper do not necessarily represent the views of the U.S. Department of Energy or the United States Government.




\clearpage
\newpage
\renewcommand{\thefigure}{S\arabic{figure}}
\setcounter{figure}{0}    
\renewcommand{\thepage}{SI-\arabic{page}}
\setcounter{page}{1}    
\begin{center}
{\bf SUPPORTING INFORMATION}
\end{center}

Here we include results on the receding speed of the liquid-vapor interface during evaporation and a direct comparison of the nanoparticle concentration profiles along the $z$-axis under the various evaporation schemes discussed in the main text.

\begin{figure}[htb]
\includegraphics[width = 0.45\textwidth]{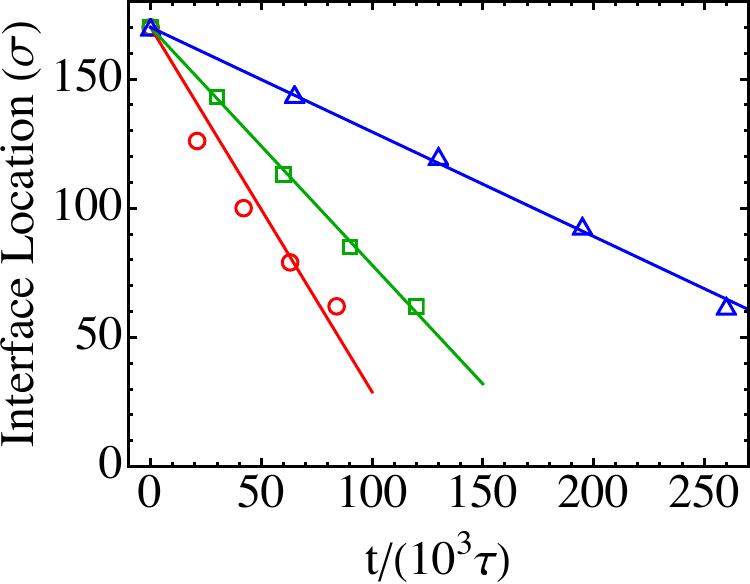}
\caption{Location of the receding liquid-vapor interface vs. time under various solvent evaporation schemes: evaporating into a vacuum (red circles), evaporating at a fixed intermediate rate of $j_p$ (the terminal plateau rate of the evaporating-into-vacuum scheme, green squares), and evaporating at a fixed rate of $j_s \simeq j_p/3$ (blue triangles). The lines are linear fits whose slopes yield the receding speed of the interface, $v_e$.}
\label{fg:interface}
\end{figure}

The location of the liquid-vapor interface during solvent evaporation is plotted against time in Fig.~\ref{fg:interface}. At a fixed evaporation rate, the liquid-vapor interface recedes almost uniformly at a constant speed, $v_e$. Under the evaporating-into-vacuum scheme, the interface location varies with time nonlinearly and the rate of change decreases as time proceeds. This reflects the fact that when the solvent vapor escapes into a vacuum, the evaporation rate is initially high, then decreases with time, and eventually levels off in the long time limit. For each data set included in Fig.~\ref{fg:interface}, a linear fit is performed to determine $v_e$. For the evaporating-into-vacuum scheme, $v_e \simeq 1.4\times 10^{-3} \sigma/\tau$. At the intermediate solvent evaporation rate, $v_e \simeq 0.92\times 10^{-3} \sigma/\tau$ and at the slow rate, $v_e \simeq 0.40\times 10^{-3} \sigma/\tau$.

\begin{figure}[htb]
\includegraphics[width = 0.45\textwidth]{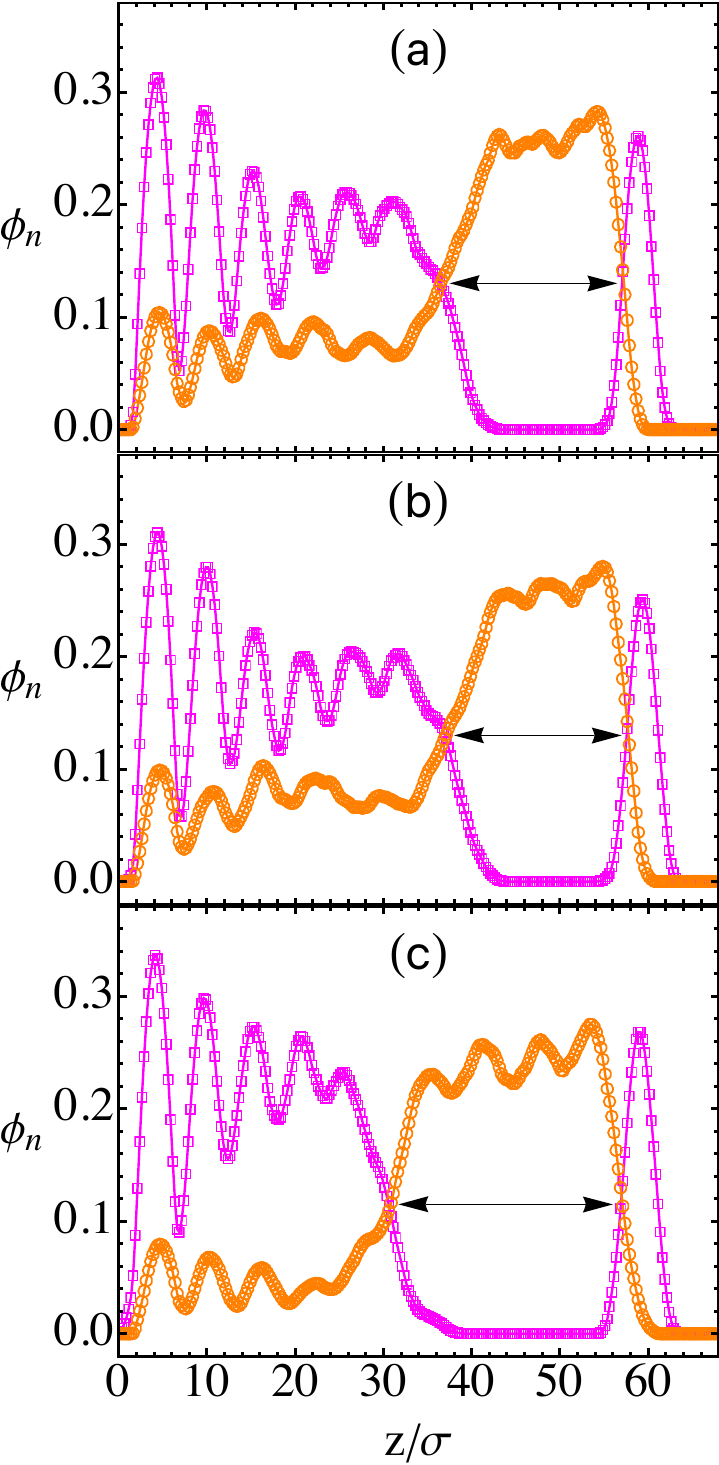}
\caption{Concentration profiles of the two types of nanoparticles ($\alpha$ in orange and $\beta$ in purple) in the drying film at a thickness of $\sim 62\sigma$ under the three solvent evaporation schemes: (a) evaporating-into-vacuum, (b) fixed intermediate rate, and (c) fixed slow rate. The arrows indicate the width of the middle $\alpha$-layer.}
\label{fg:final_np_dist}
\end{figure}

The concentration profiles of $\alpha$ and $\beta$ nanoparticles in the drying film of a thickness of about $62\sigma$ are compared directly in Fig.~\ref{fg:final_np_dist}. Under the slow evaporation rate, the two types of nanoparticles phase separate more strongly and the middle $\alpha$-layer is thicker. The concentration of $\alpha$ nanoparticles in the bottom $\beta$-layer is also smaller in this case than under the other two evaporation schemes.

To quantify the thickness of the middle $\alpha$-layer, we identify the crossing points in the concentration profiles, as indicated by the arrows in Fig.~\ref{fg:final_np_dist}. The thickness is computed as the difference of the two crossing points, on the two sides of the $\alpha$-layer, along the $z$-axis. It is about $21\sigma$ for the evaporating-into-vacuum and intermediate-rate schemes and about $26\sigma$ for the slow-rate scheme.

\end{document}